\documentclass{article}
\usepackage{spconf,amsmath,epsfig}
\usepackage{amsfonts}
\usepackage{amsmath}
\usepackage{mathrsfs}
\usepackage{multirow}
\usepackage{subfigure}

\begin{document}\sloppy

\def\x{{\mathbf x}}
\def\L{{\cal L}}

\title{Learned Scalable Image Compression with Bidirectional Context Disentanglement Network}
%
\name{Zhizheng Zhang \qquad Zhibo Chen$^*$\thanks{$^*$Corresponding author. (E-mail: {\tt\small chenzhibo@ustc.edu.cn})} \qquad Jianxin Lin \qquad Weiping Li}
\address{CAS Key Laboratory of Technology in Geo-spatial Information Processing and Application System,\\ University of Science and Technology of China, Hefei, China}

\maketitle

\begin{abstract}
In this paper, we propose a learned scalable/progressive image compression scheme based on deep neural networks (DNN), named Bidirectional Context Disentanglement Network (BCD-Net). For learning hierarchical representations, we first adopt bit-plane decomposition to decompose the information coarsely before the deep-learning-based transformation. However, the information carried by different bit-planes is not only unequal in entropy but also of different importance for reconstruction. We thus take the hidden features corresponding to different bit-planes as the context and design a network topology with bidirectional flows to disentangle the contextual information for more effective compressed representations. Our proposed scheme enables us to obtain the compressed codes with scalable rates via a one-pass encoding-decoding. Experiment results demonstrate that our proposed model outperforms the state-of-the-art DNN-based scalable image compression methods in both PSNR and MS-SSIM metrics. In addition, our proposed model achieves better performance in MS-SSIM metric than conventional scalable image codecs. Effectiveness of our technical components is also verified through sufficient ablation experiments.
\end{abstract}
\begin{keywords}
scalable image compression, deep learning
\end{keywords}
\section{Introduction}
\label{sec:intro}

Scalable coding is different from non-scalable coding in the sense that the coded bitstream of scalable coding is partially decodable. That is to say that scalable image compression allows reconstructing complete images with more than one quality levels simultaneously by decoding appropriate subsets of the whole bitstream, which is called the ``bitstream scalability''. In terms of the comparison with a simulcast codec \cite{mccanne1996scalable}, a scalable codec produces a cumulative set of hierarchical representations which can be combined for progressive refinement, instead of producing a multi-rate set of signals that are independent of each other. Thereby, scalable/progressive image compression is of great significance for image transmission and storage in practical using.

As the prosperity of deep learning, DNN-based models for lossy image compression \cite{toderici2015variable,balle2016end,toderici2017full,theis2017lossy,agustsson2017soft,baig2017learning,balle2018variational,minnen2018joint} have been widely explored recently. Toderici et al. \cite{toderici2017full} and Baig et al. \cite{baig2017learning} study the design of network architectures for deep image compression. Ball$\acute{e}$ et al. \cite{balle2016end} and Agustsson et al. \cite{agustsson2017soft} introduce the trainable quantization methods to help achieving end-to-end optimization. The works of \cite{balle2016end,baig2017learning,rippel2017real,toderici2017full,mentzer2018conditional} investigate the context models to improve the compression efficiency of arithmetic coding. In addition, the biologically-inspired joint nonlinearity, named generalized divisive normalization (GDN), is proposed in \cite{balle2016end}. And the structure of side information in learned image compression is well studied in \cite{balle2018variational, minnen2018joint}. Specially, Agustsson et al.\cite{agustsson2018generative} employ the generative models in improving the perceptual performance of learned image compression. Scalabilty, as a critical property, has not drawn much attention explicitly in deep-learning-based schemes, while it is supported by many prevailing conventional image/video compression standards \cite{skodras2001jpeg,li2001overview,ye2014scalable}.

In terms of the related works based on deep learning, Gregor et al. \cite{gregor2016towards} introduce a novel hierarchical representation of images with a homogeneous deep generative model, which is considered as a "conceptual compression" framework instead of a real compressor. The framework proposed by Toderici et al. \cite{toderici2017full} can be viewed as the first DNN-based image compression model supporting bitstream scalability, in which recurrent neural networks (RNN) are employed to compress the residual information of the last reconstruction relative to the original image iteratively. However, \cite{toderici2017full} still suffers from limited rate-distortion performance and complex encoding-decoding process due to the multi-iteration encoding-decoding within it.

In this paper, we are devoted to developing a more effective learned scalable image compression scheme. Except for obtaining better rate-distortion performance, we also aim to develop the functionality of enabling us to obtain the reconstructed images with different quality levels via a one-pass encoding-decoding simultaneously. Inspired by the Fine Granularity Scalability (FGS) \cite{li2001overview} in MPEG-4 video standard, we adopt bit-plane decomposition to decompose the information before the input layer of neural networks. Bit-plane decomposition has an inherent advantage to transform the image to a hierarchical representation, in which an RGB image can be transformed into 24 bit-planes losslessly (8 bit-planes per channel). Two significant things can be observed: firstly, the sum of the information entropy \cite{wyner1974recent} (shortly called ``entropy'') of all bit-planes always exceeds the entropy of the corresponding original image; secondly, different bit-planes are not equal in their entropy. Theoretically, the carried information of a particular sequence of independent events is the sum of the information carried by each event. Therefore, there should be a correlation among different bit-planes, which is hard to be well considered in conventional bit-plane coding. In addition, the information carried by different bit-planes are asymmetrical due to their unequal entropy volumes. In this work, we make the first endeavour to employ deep neural networks in capturing the correlation among bit-planes in coding process. Moreover, for the information with different importance for reconstruction, we design a self-consistent architecture to disentangle them to form the hierarchical representations with an end-to-end optimization.

In summary, we have made three main contributions: (1) We propose a new DNN-based framework for learned scalable/progressive image compression, which can enable us to get the compressed results corresponding to multiple bitrates simultaneously through one-pass encoding and decoding. Note that the only one previous DNN-based image codec \cite{toderici2017full} can support bitstream scalability and it requires multi-iteration encoding and decoding to get compressed results with different quality levels. (2) We propose to involve the idea of bit-plane coding into a learnable scalable image codec, which benefits informantion decomposition for more effective hierarchical representation. (3) Within our proposed model, we design a LSTM-based architecture to disentangle the information of different bit-planes and achieve an end-to-end optimization for better rate-distortion performance, which goes beyond the regular using of LSTMs \cite{hochreiter1997long}. Our proposed method outperforms the state-of-the-art DNN-based scalable image codec greartly in both PSNR and MS-SSIM metrics.

\section{Proposed Method}


We propose a deep-learning-based framework for scalable/progressive image compression. Within this framework, we adopt bit-plane decomposition to perform information decomposition coarsely and design two bidirectional gated units to disentangle the contextual information precisely.

\begin{figure*}[t]
\centerline{\includegraphics[scale=0.58]{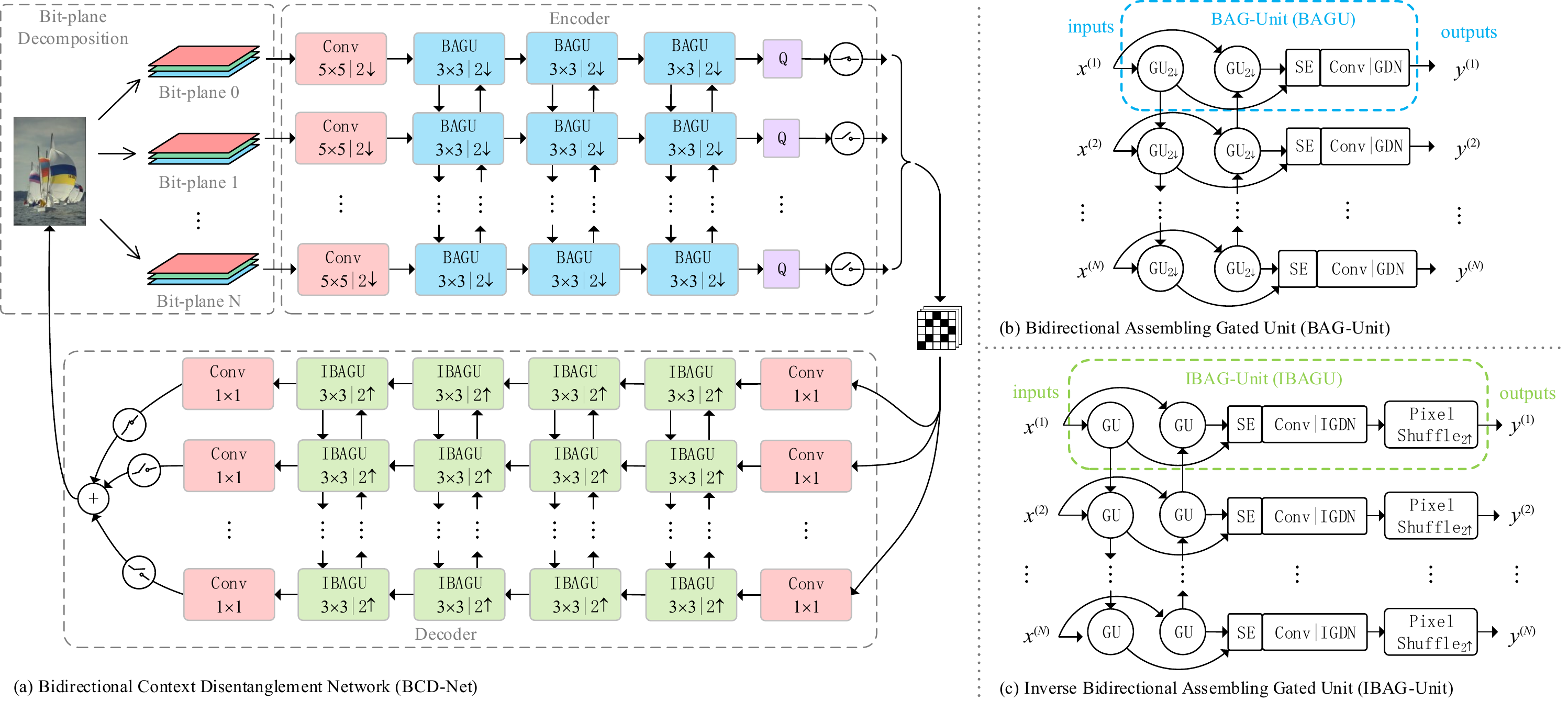}}
\caption{Network structures of (a) the Bidirectional Context Disentanglement Network (BCD-Net) which includes bit-plane decomposition, an image encoder and an image decoder, (b) the Bidirectional Assembling Gated Unit (BAG-Unit, or BAGU), (c) the Inverse Bidirectional Assembling Gated Unit (IBAG-Unit, or IBAGU). In (a), the ``$Q$'' denotes the quantization module to transfer the feature map into a binary tensor, which is the same as the ``Binarizer'' defined in \cite{toderici2017full}. The switch symbol represents an our defined operation that is described detailedly in the text. The $x\times y$ indicates the size of convolution kernels in the corresponding units, and the symbol ``+'' denotes the element-wise adding operation. In (b) and (c), the ``GU''is short for ``Gated Unit'', and the ``SE'' is short for ``Squeeze-and-Excitation'' proposed in \cite{hu2017squeeze}. The ``GDN'' is the generalized divisive normalization proposed in \cite{balle2016end}, and the ``IGDN'' refers to the ``Inverse-GDN'' that is also defined in \cite{balle2016end}. The ``Pixel Shuffle'' refers to a depth-to-space operation to perform up-sampling. In all of the sub-figures, ``$2\downarrow$'' represents ``$2\times$'' down-sampling for the input hidden variables, while ``$2\uparrow$'' represents ``$2\times$'' up-sampling for the input hidden variables.}
 \label{fig:scheme}
\end{figure*}

\subsection{Scalable Compression Framework}

\textbf{Bit-plane decomposition.} As illustrated in Fig.\ref{fig:scheme}.(a), for a RGB image, we transfer each channel of it into $N$ bit-planes through bit-plane decomposition. Here $N$ can be viewed as so-called bit-depth. In this paper, we set $N=8$ for the RGB images in which the pixels are in the range of [0, 255]. For clarity, we represent the $l^{th}$ bit-plane for R, G, B channels as $X_{R}^{(l)}$, $X_{G}^{(l)}$, $X_{B}^{(l)}$ respectively. Illustratively, we denote the pixel located at $(h, w)$ in $R$ channel as $X_R(h,w)$, then we can obtain its corresponding value in the $l^{th}$ bit-plane as below:
\begin{equation}\label{eq:1}
X_{R}^{(l)}(h,w) = \begin{cases}
1, & \text{if } \lfloor X_{R}(h,w)/2^{N-l}\rfloor\,mod\,2=1 \\
0, & \text{otherwise}
\end{cases}
\end{equation}
where $\left \lfloor x  \right \rfloor$ is the function to get the greatest integer less than or equal to $x$. Inversely, we can reconstruct the original image from bit-planes by the following formula:
\begin{equation}\label{eq:2}
X_{R}(h,w)=\sum_{l=1}^{N}2^{N-l}\times X_{R}^{(l)}(h,w).
\end{equation}
By the operation described in Eq.\ref{eq:1}, the original information from the RGB images is unevenly scattered into eight correlated but heterogeneous sub-spaces. However, Eq.\ref{eq:2} shows that each bit-plane is of different importance for reconstruction. In addition, since the information entropy of each bit-plane is not equal, the information volume carried by each bit-plane is also different.

\textbf{Encoder.} Taking the bit-planes as the input of the encoder, we design a multi-branch architecture to learn the hierarchical representations. The network layers in each branch don't share weights with the layers in other branches. As shown in Fig.\ref{fig:scheme}.(a), we leverage one convolutional layer to perform a preliminary transformation for each bit-plane independently, followed by three layers consisting of BAG-Units to further transform the carried information and yield $N$ feature map partitions. Notice that there are bidirectional information flows between the two BAG-Units in adjacent branches via their hidden states. In both of the first convolutional layers and BAG-Units, we use the convolutional operation with the stride of 2 to achieve spatial down-sampling for the feature maps. The quantization module ``$Q$'' includes a $1\times1$ convolutional layer with the stride of 1, a tanh activation and a binarization function defined in \cite{toderici2015variable}. In the final of the encoder, we define a switch function located in each branch. When the switch is ``on'', the corresponding feature map will be retained as one participant of the compressed codes before entropy coding; when the switch is ``off'', the corresponding feature map will be filled with zero values in the compressed codes before entropy coding. Finally, we employ the method of entropy coding in \cite{toderici2017full} for the codes from each branch individually to obtain the final compressed codes.

In terms of the transmission, only the compressed codes corresponding to the switch in ``on'' state should be transmitted from the sender to the receiver. The sum of the rates of all final compressed codes determines the compression rate and the highest reconstructed quality level in our scalable framework. The ``basic bitrate'', namely the minimal coding rate can be achieved for one trained model before entropy coding, depends on the size of the binary feature map after quantization per branch in the encoder network. Some related recommended settings are listed in our supplementary materials.

\textbf{Decoder.} In our decoder, we leverage a $1\times1$ convolutional layer with the stride of 1 to tune the dimensions of feature maps at the beginning and the ending of decoding respectively. Then we also use a multi-branch architecture to disentangle the contextual information for reconstruction in the procedure of decoding. Different from the method of down-sampling in the encoder, here we use pixel shuffle such a depth-to-spatial operation to implement spatial up-sampling. The same switch function in our decoder is used for controlling the quality level of the reconstructed image.

\subsection{Contextual Information Disentanglement}

In this section, we elaborate the architecture design for BAG-Unit and IBAG-Unit which play the role of disentangling contextual information in our scalable compression framework. We design a modified version of bidirectional-convolutional LSTM \cite{liu2017bidirectional} as the gated unit in BAG-Unit and IBAG-Unit and go beyond the regular using of LSTMs.

The information of each bit-plane is heterogeneous with each other. We therefore propose to abandon the recurrent connections of LSTM units by using different units with unshared weights. Mathematically, let $x^{(l)}$, $c^{(l)}$, $h^{(l)}$ and $y^{(l)}$ denote the input, cell, hidden, and output states of the BAG-Unit/IBAG-Unit in the $l^{th}$ branch (see (b) and/or (c) in Fig.\ref{fig:scheme}). Clearly, we use the arrows above the symbol to distinguish two different directions of the information flows between adjacent BAG-Units/IBAG-Units. For the paired gated unit within the BAG-Unit/IBAG-Unit in the $l^{th}$ branch, their cell, hidden and output states can be updated as follows:

\begin{equation}\label{eq:3.1}
\begin{aligned}
\overrightarrow{i}^{(l)}&=\sigma(\overrightarrow{W}_{ix}^{(l)}*x^{(l)}+\overrightarrow{W}_{ih}^{(l)}*h^{(l-1)}+\overrightarrow{b}_{i}^{(l)}),\\
\overleftarrow{i}^{(l)}&=\sigma(\overleftarrow{W}_{ix}^{(l)}*x^{(l)}+\overleftarrow{W}_{ih}^{(l)}*h^{(l+1)}+\overleftarrow{b}_{i}^{(l)}),
\end{aligned}
\end{equation}

\begin{equation}\label{eq:4}
\begin{aligned}
\overrightarrow{f}^{(l)}&=\sigma(\overrightarrow{W}_{fx}^{(l)}*x^{(l)}+\overrightarrow{W}_{fh}^{(l)}*h^{(l-1)}+\overrightarrow{b}_{f}^{(l)}),\\
\overleftarrow{f}^{(l)}&=\sigma(\overleftarrow{W}_{fx}^{(l)}*x^{(l)}+\overleftarrow{W}_{fh}^{(l)}*h^{(l+1)}+\overleftarrow{b}_{f}^{(l)}),
\end{aligned}
\end{equation}

\begin{equation}\label{eq:5}
\begin{aligned}
\overrightarrow{o}^{(l)}&=\sigma(\overrightarrow{W}_{ox}^{(l)}*x^{(l)}+\overrightarrow{W}_{oh}^{(l)}*h^{(l-1)}+\overrightarrow{b}_{o}^{(l)}),\\
\overleftarrow{o}^{(l)}&=\sigma(\overleftarrow{W}_{ox}^{(l)}*x^{(l)}+\overleftarrow{W}_{oh}^{(l)}*h^{(l+1)}+\overleftarrow{b}_{o}^{(l)}),
\end{aligned}
\end{equation}

\begin{equation}\label{eq:6}
\begin{aligned}
\overrightarrow{in}^{(l)}&=\tanh(\overrightarrow{W}_{inx}^{(l)}*x^{(l)}+\overrightarrow{W}_{inh}^{(l)}*h^{(l-1)}+\overrightarrow{b}_{in}^{(l)}),\\
\overleftarrow{in}^{(l)}&=\tanh(\overleftarrow{W}_{inx}^{(l)}*x^{(l)}+\overleftarrow{W}_{inh}^{(l)}*h^{(l+1)}+\overleftarrow{b}_{in}^{(l)}),
\end{aligned}
\end{equation}

\begin{equation}\label{eq:7}
\begin{aligned}
\overrightarrow{c}^{(l)}&=\overrightarrow{f}^{(l)}\odot\overrightarrow{c}^{(l-1)}+\overrightarrow{i}^{(l)}\odot\overrightarrow{in}^{(l)},\\
\overleftarrow{c}^{(l)}&=\overleftarrow{f}^{(l)}\odot\overleftarrow{c}^{(l+1)}+\overleftarrow{i}^{(l)}\odot\overleftarrow{in}^{(l)},
\end{aligned}
\end{equation}

\begin{equation}\label{eq:8}
\overrightarrow{h}^{(l)}=\overrightarrow{o}^{(l)}\odot\tanh(\overrightarrow{c}^{(l)}),\overleftarrow{h}^{(l)}=\overleftarrow{o}^{(l)}\odot\tanh(\overleftarrow{c}^{(l)}).
\end{equation}
where ``$*$'' denotes the convolutional operator, and ``$\odot$'' denotes element-wise multiplication. The symbols $i$, $f$ and $o$ represent the input gate, forget gate and output gate respectively, and $in$ indicates the input of the gated unit. Additionally, $W$ with different subscripts denote the weight matrices of different convolutional transformations, and $\sigma$ denotes the sigmoid activation function $\sigma(x)=1/(1+exp(-x))$. The output state $y^{(l)}$, which is also the input of ``SE'' block, is the result of concatenating the hidden states $\overrightarrow{h}^{(l)}$ and $\overleftarrow{h}^{(l)}$ of the gated units in two directions. 

The gated units in BAG-Unit/IBAG-Unit play two important roles in disentangling the information: (1) capturing the correlations among different bit-planes, which benefits reducing rate for compact representations in compression; (2) helping to determine which level of feature partitions the information should be expressed according to its relative importance.  

After the gated units, we employ the ``Squeeze-and-Excitation'' module to introduce a channel-wise attention for better fusing the information from different directions. Then we use a $3\times3$ convolution layer with the stride of 1 to perform further transformation.

\subsection{Training Algorithm}

As a scalable image compression framework, it is required to be optimized for hierarchical reconstructed results with different quality levels meanwhile during training. Therefore, we use a specific approach to train this model, in which each training step contains a one-pass forward process of the encoder, a multi-pass forward process of the decoder and a one-pass backward process for parameters updating. Clearly, suppose that there are $N$ quality levels in all, the loss function can be depicted by the formula below:
\begin{equation}\label{eq:10}
Loss=\sum_{l=1}^{N}\beta^{(l)}\cdot\mathbb{D}(\hat{X}^{(l)}, X),\quad \hat{X}^{(l)}=\sum_{i=1}^{l}Y^{(i)}
\end{equation}
where $\hat{X}^{(l)}$ denotes the reconstructed results at the level of $l$, $Y^{(i)}$ represents the output of the i-th branch, and $\mathbb{D}(\cdot)$ refers to the distance function which is related to the distortion metrics used for evaluation. Here, we take L1 norm and MS-SSIM (proposed in \cite{wang2003multiscale}) as the mentioned distance functions to train our model in this paper. We weight the distortions under different code rates with a coefficient $\beta^{(l)}$, which is set to $1/N$ generally.

\section{Experiment Results}
\subsection{Datasets and Settings}

We use two sets of training data to train our proposed model, which includes the CoCo dataset \cite{lin2014microsoft} and a dataset composed of thirty thousand RGB images we collected from the word wide web. For the first dataset, we obtain $n\times n$ ($n$ can be taken as 32, 64 and 128) image patches for training by adopting the commonly used data augmentation strategies of random cropping and random horizontal flipping (with a probability of 0.5). For the second dataset, each image is first scaled by a random factor in [0.5, 1.5], followed by a random cropping and a random horizontal flipping (with a probability of 0.5). Then, we perform filtering the obtained image patches by using the sobel operator and cany operator to reduce the ratio of the training samples with too simple textures.

We implement three-stage training procedures with different patch sizes at each stage for our proposed models. We first pre-train our model by using $32\times 32$ patches from the first dataset and perform stochastic gradient descent with minibatches of 32 by adopting Adam optimizer with a learning rate of $5\times 10^{-5}$. Then we train our model by using $64\times 64$ patches from the second dataset. At this stage, we set the size of minibatch as 32 and adopt Adam optimizer with a initial learning rate of $5\times 10^{-4}$ and a weight decay of $5\times 10^{-4}$. We finally perform fine-tuning with $128\times 128$ image patches from the second dataset. At this stage, we tune $\beta^{(l)}$ of Eq.10 in main text in a small range for improving the performance with respect to some specific bitrates.

\begin{figure*}[t]
\subfigure
{\includegraphics[width=.96\columnwidth]{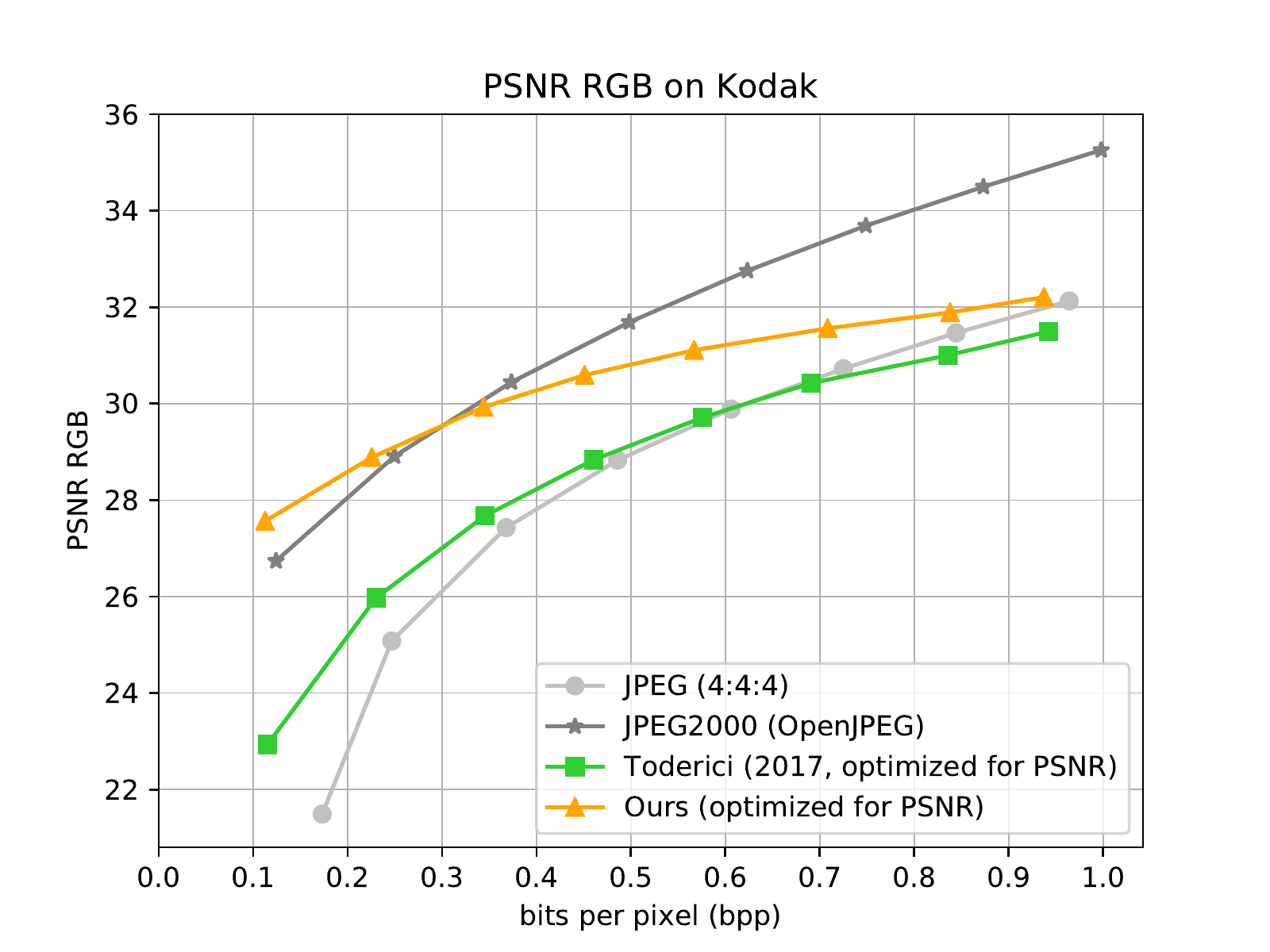}}
\hspace{.38cm}
\subfigure
{\includegraphics[width=.96\columnwidth]{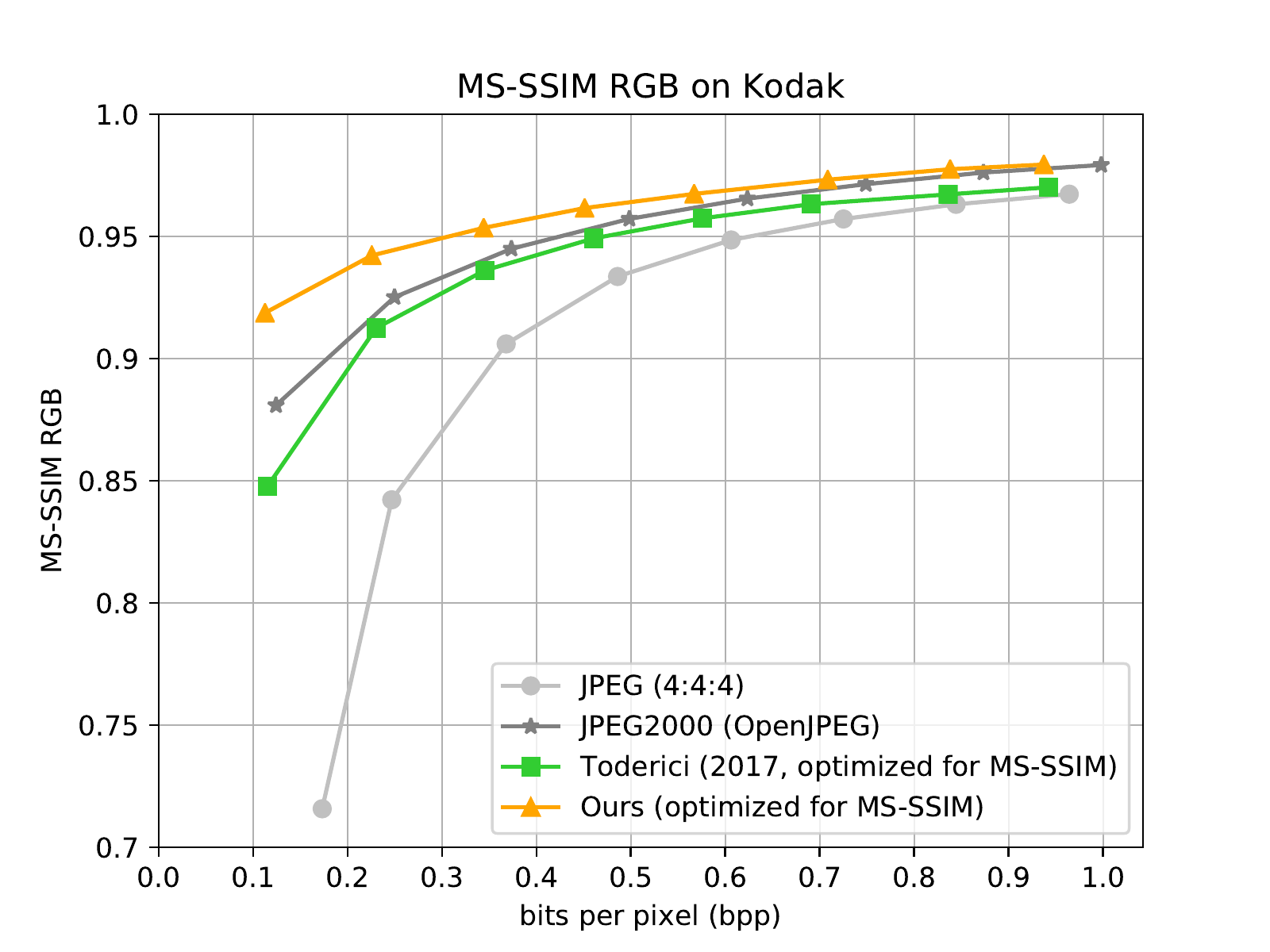}}
\caption{Rate-distortion performance evaluated on the Kodak dataset. The left figure is given as PSNR vs. bits per pixel (bpp), while the right figure is given as MS-SSIM vs. bpp. The reported results of Toderici et al. \cite{toderici2017full} are obtained from our reproduced model, which is very close to their original reported results. To enable a fair comparison, we use the same training data with ours and adopt the training settings in their publication \cite{toderici2017full}. Note that the JPEG and JPEG2000 are the conventional image codecs.}
\label{fig:rate_distortion}
\end{figure*}
    


\subsection{Rate-distortion Performance}

We evaluate our proposed models on the Kodak dataset and illustrate the best rate-distortion performance across multiple trained models under different bitrates in Fig.\ref{fig:rate_distortion}. By involving bit-plane decomposition and disentangling the information with the BCD-Net, in both PSNR and MS-SSIM metrics, our proposed model achieves a significant improvement across different bit-rates compared to the current state-of-the-art DNN-based scalable image compression model. Relative to the conventional scalable image codec JPEG2000, our proposed model outperforms it in MS-SSIM metric across different bit-rates, and it also shows its advantage in PSNR metric under low bit-rates. 

\subsection{Ablation Study}

\begin{table*}[t]
\centering
\newcommand{\tabincell}[2]{\begin{tabular}{@{}#1@{}}#2\end{tabular}}
\caption{The numerical evaluation results of ablation study.}
\label{ablation}
\footnotesize
\resizebox{\textwidth}{!}{%
\begin{tabular}{|c|c|c|c|c|c|c|c|c|c|}
\hline
\multicolumn{2}{|c|}{\multirow{2}{*}{\begin{tabular}[c]{@{}c@{}}Rate (bpp) \&\\ Distortion\end{tabular}}} & \multicolumn{2}{c|}{0.0625} & \multicolumn{2}{c|}{0.1250} & \multicolumn{2}{c|}{0.1875} & \multicolumn{2}{c|}{0.2500} \\ \cline{3-10} 
\multicolumn{2}{|c|}{} & PSNR & MS-SSIM & PSNR & MS-SSIM & PSNR & MS-SSIM & PSNR & MS-SSIM \\ \hline
\multirow{4}{*}{\tabincell{c}{(1) Unidirectional \\ Encoder-decoder}} & $E\downarrow D\downarrow$ & 22.6267 & 0.7630 & 25.3592 & 0.8448 & 27.1178 & 0.8840 & 27.7594 & 0.9016 \\ \cline{2-10} 
 & $E\downarrow D\uparrow$ & 22.6432 & 0.7684 & 25.5035 & 0.8536 & 26.7740 & 0.8830 & 27.4419 & 0.8966 \\ \cline{2-10} 
 & $E\uparrow D\downarrow$ & 24.7268 & 0.7863 & 25.5717 & 0.8183 & 25.6212 & 0.8194 & 25.6195 & 0.8194 \\ \cline{2-10} 
 & $E\uparrow D\uparrow$ & 24.5063 & 0.7894 & 25.5561 & 0.8216 & 25.6693 & 0.8240 & 25.6720 & 0.8244 \\ \hline
\multicolumn{2}{|l|}{(2) with the regular using of LSTMs} & 25.3584 & 0.8175 & 26.5429 & 0.8707 & 27.2956 & 0.8986 & 27.4378 & 0.9030 \\ \hline
\multicolumn{2}{|l|}{(3) w/o bit-plane decomposition} & 25.1074 & 0.8203 & 26.5585 & 0.8720 & 27.2947 & 0.8929 & 27.6120 & 0.9036 \\ \hline
\multicolumn{2}{|l|}{(4) w/o the ``SE'' modules} & 25.6104 & 0.8163 & 26.8601 & 0.8721 & 27.5019 & 0.8948 & 27.8937 & 0.9051 \\ \hline
\multicolumn{2}{|l|}{(5) w/o the GDN/IGDN} & 25.3693 & 0.8160 & 26.6676 & 0.8715 & 27.3263 & 0.8932 & 27.7023 & 0.9027 \\ \hline
\multicolumn{2}{|l|}{(6) Fully-equipped BCD-Net} & \textbf{25.8295} & \textbf{0.8297} & \textbf{27.304}5 & \textbf{0.8785} & \textbf{27.9695} & \textbf{0.8999} & \textbf{28.3327} & \textbf{0.9101} \\ \hline
\end{tabular}}
\end{table*}
To further investigate the effectiveness of the technical components within our proposed scheme, we construct a series of experiments to compare our proposed BCD-Net with the following individual experimental cases: (1) We implement four different combinations of an unidirectional encoder and an unidirectional decoder, in which ``E'' and ``D'' denote the encoder and the decoder respectively, and the symbols $\downarrow$ and $\uparrow$ represent two directions of the information flow; (2) We take the LSTM with recurrent connections as the gated units inside BAG-Units and IBAG-Units; (3) We replace the bit-plane decomposition with convolution and slicing operations; (4) We take the SE blocks away from BAG-Units and IBAG-Units. (5) We replace the GDN and IGDN inside BAG-Units and IBAG-Units with the non-linear activation function Leaky ReLU ($\alpha = 0.2$).

For each experimental case of our above descriptions, we train the corresponding model with the basic bitrate of $1/32$ bits per pixel (bpp) and keep the settings other than its individual described factor same with \emph{Fully-equipped BCD-Net} for fair comparisons. All experimental cases here are optimized for PSNR metric under the same training settings. The evaluation results on the Kodak dataset are reported in Table.\ref{ablation}.

As shown in Table.\ref{ablation}, the rate-distortion performance of codecs decline severely when we apply the unidirectional network topology in encoder and decoder, which shows that the bidirectional information flow is crucial for context disentanglement. Bidirectional message passing helps for determining where should be expressed for the information with different importance and considering the correlations among the representations at different levels. The second experiment demonstrates that the LSTMs with unshared parameters are more suitable for mapping heterogeneous information hidden in different subspaces to latent representations when compared to the regular using of LSTMs with recurrent connections. Also, we can find that bit-plane decomposition is better than convolution and slicing operations in providing a coarse but effective information decomposition before the deep-learning-based transformation. The results of ablation study also suggest that ``Squeeze-and-Excitation'' block can lead to better information fusion by introducing the channel-wise attention. Additionally, similar with Ball$\acute{e}$ et al.'s work \cite{balle2016end}\cite{balle2018variational}, GDN/IGDN is also effective within our scheme in simplifying learning by Gaussianizing image densities. The future works include more effective entropy coding and post-processing for DNN-based scalable image codecs.

\section{Conclusions}

In this paper, we study the deep-learning-based scalable image codec. We propose to involve bit-plane decomposition in a DNN-based compression framework to decompose the original information coarsely. Then we design the Bidirectional Context Disentanglement Network (BCD-Net) to learn more effective hierarchical representations for scalable/progressive compression. Consequently, our proposed model can compress and reconstruct the images with different quality levels simultaneously through a one-pass encoding-decoding. And it outperforms the state-of-the-art of DNN-based scalable image codecs in both PSNR and MS-SSIM metrics. It also outperforms the conventional scalable image codec in MS-SSIM metric across different bitrates and in PSNR metric under low bitrates.

\section{Acknowledgement}
This work was supported by the National Key Research and Development Program of China under Grant No. 2016YFC0801001, the National Program on Key Basic Research Projects (973 Program) under Grant 2015CB351803, NSFC under Grant 61571413, 61390514.

\small
\bibliographystyle{IEEEbib}
\bibliography{main}

\end{document}